**Absolutely Zero Evidence**

Veronica J. Vieland

> Battelle Center for Mathematical Medicine
> Departments of Pediatrics and Statistics
> The Research Institute at Nationwide Children's Hospital
> & The Ohio State University

Address corresondence to: veronica.vieland@nationwidechildrens.org

**Abstract**  Statistical analysis is often used to evaluate the evidence for or against scientific hypotheses, and various statistics (e.g., p-values, likelihood ratios, Bayes factors) are interpreted as measures of evidence strength. Here I consider evidence measurement from the point of view of representational measurement theory, and argue that familiar evidence statistics do not conform to any legitimate measurement scale type. I then consider the notion of an *absolute* scale for evidence measurement, in a sense to be defined, focusing particularly on the notion of absolute 0 evidence, which turns out to be something other than what one might have expected.

## 1. Introduction

Statistical analysis is common in the biological and social sciences, and certain statistical outputs are routinely understood in evidential terms. The most commonly used evidence statistic (ES) is the empirical p-value (P); other familiar ESs include the maximum likelihood ratio (MLR) and Bayes factor (BF)[1]. Small values of P are routinely said to indicate strong *evidence* against a null hypothesis, with *evidence strength* taken to be stronger the smaller the value of P. In the context of replication, failure to achieve a sufficiently small P on follow up is interpreted as indicating *evidence* against an initial finding. And scientific studies with results reported in terms of P are often summarized by the media with "study finds *evidence* of…".

One could argue that interpreting P in evidential terms is a mistake. What seems unarguable, though, is that statistical evidence is something scientists often *want* to measure, and it is something that scientists *do* measure, whenever they treat the numerical value of an ES as representing evidence strength. While we don't usually think of it this way, this constitutes an act of measurement. Here I apply basic precepts of representational measurement theory (Hand 2004) in considering the measurement of statistical evidence.

---

[1] P is the probability of obtaining data (or test statistic T) at least as extreme as the observed data (or T), assuming the null hypothesis $H_0$ is true. MLR is the likelihood for the observed data assuming one hypothesis ($H_1$) divided by the likelihood for those same data assuming another hypothesis ($H_2$), where free parameters (i.e., parameters with unspecified values) are eliminated via maximization. For convenience I assume nested hypotheses (same parameters in numerator and denominator) with the number of free parameters in the numerator $\geq$ the number in the denominator. BF is the same as MLR, except that free parameters are eliminated via weighted averaging, which entails introducing prior probability distributions.

The paper is organized as follows. In (2) I review distinctions among the three major types of measurement scales (ordinal, interval and ratio), then in (3) I consider the scale types of different ESs. (Spoiler alert: familiar ESs cannot be coherently assigned to *any* valid scale type.) In (4) I consider the absolute scale, a type of ratio scale usually only discussed in connection with temperature, focusing in particular on what makes the 0-point of some but not all ratio scales *absolute*. In (5) I explore 0-points for ESs. This leads, in (6), to a counter-intuitive conclusion regarding absolute 0 evidence.

## 2. Measurement Scale Types

Measurements can be classified into different scale types, which can be characterized in (at least) two useful ways. The first is by asking what we can do *with* them. To illustrate, suppose test subjects are asked to make a judgment of "more beautiful" or "less beautiful" in a series of pair-wise comparisons among $n$ pictures of different faces. This allows a rank-ordering of the pictures from least beautiful to most, and we can assign numbers (say, 1 to $n$) to represent this ordering. We can now compare faces with respect to rank-ordering, e.g., we can say "face #100 is judged to be more beautiful than #99." However, it is *not* meaningful to ask whether the amount by which #100 is more beautiful than #99 is greater than the amount by which #50 is more beautiful than #49, or than #2 for that matter. This is because nothing in the way we have constructed the scale allows us to interpret distances from one number to another in terms of underlying units of beauty. This illustrates ordinal measurement. What we can do *with* ordinal scales is to make comparisons regarding order, and nothing more.

A second way to characterize scale types is by asking what we can do *to* them. For ordinal variables, we can transform the original scale into any other set of symbols that preserves rank-ordering, e.g., we can replace our scale values $1,..,n$ with their respective logarithms. As long as the transformation preserves rank-order, it preserves the meaning of the original scale.

The logarithmic transformation would no longer be meaning-preserving, however, if we interpreted the difference between numbers on the original scale as having meaningful units. For instance, the Fahrenheit temperature scale provides not only a rank-ordering of temperatures, but also assigns meaning to the unit: the difference in temperature between 100°F and 99°F is the same as the difference in temperature between 50°F and 49°F. We can express this by saying that 1°F always "means the same"[2] with respect to temperature. This illustrates an interval scale. What we can do with interval scales is to meaningfully make comparisons of order and also of differences. As for what we can do to them, interval scales are amenable to any linear transformation (e.g., the formula converting °F to °C), because such transformations preserve both rank-order and a constant meaning for the unit across the scale range. The principal difference between °F and °C is in the size of the degree, or the *thermal meaning* of one degree.[3]

---

[2] This is Hacking's (1972, 136) phrase, from a passage critiquing the LR as an ES because no argument exists to show that a unit change in the LR always "means the same."

A logarithmic transformation would disrupt this thermal meaning. $\text{Log}_e(100°F)-\log_e(99°F)=0.01$ while $\log_e(50°F)-\log_e(49°F)=0.02$. Thus the same change in temperature (1°F) becomes represented by different numbers on the logarithmic scale. Application of a non-linear transformation to measurements made on an interval scale results in a "rubber scale," for which the meaning of the unit changes across the range of the scale (Houle 2011). Clearly, comparing differences on a rubber scale is problematic, in much the same way that comparing differences on an ordinal scale is problematic.

Ratio scales are interval scales with one additional feature: they count up from 0. Virtually all fundamental measurements in the hard sciences are on ratio scales, including length, weight, mass, etc. Measurements made on ratio scales can be compared with respect to order, differences, and also, ratios. For instance, temperature in degrees Kelvin (°K) has ratio scale type, which means that 20°K is twice as hot as 10°K (a ratio comparison). By contrast, such comparisons cannot be meaningfully made on the Fahrenheit (interval) scale; 20°F cannot be meaningfully said to be twice as hot as 10°F. The only arbitrary feature of a ratio scale is the size of the degree, that is, the amount of change in the object of measurement that we choose to assign to a one unit change on the measurement scale. Thus for ratio scales, the only transformation that is meaning-preserving is multiplication by a positive constant. Table 1 (modified from (Houle 2011)) summarizes the major points of this section.

---

[3] That is, the difference in temperature between 100°F and 99°F (=1°F) is *not* the same as the difference in temperature between 100°C and 99°C (=1.8°F), but the difference in temperature between 100°C and 99°C and between 50°C and 49°C remains the same (viz., 1°C, or 1.8°F). The two scales are also displaced relative to the freezing point of water (assigned 0°C and 32°F).

Ideally, we would like to be able to say when one study's evidence is twice as strong as another's. Thus I assume we would prefer, if possible, to measure evidence on a ratio scale.

**Table 1 Overview of Measurement Scale Types**

| Scale Type | Domain | Examples | Meaningful Comparisons | Permissible Transformations |
|---|---|---|---|---|
| **Ordinal** | ordered symbols | personal preference | order | rank-order preserving |
| **Interval** | real numbers | dates; temperature in °F or °C | order, differences | linear |
| **Ratio** | positive real numbers | length; mass; temperature in °K | order, differences, ratios | multiplication by a positive constant |

## 3. ES Scale Types

What can we say about the scale types of various ESs, e.g., P? When we interpret smaller values of P as stronger evidence against the null hypothesis ($H_0$), we are treating P as providing a rank-ordering of evidence, i.e., as being at least ordinally scaled. In fact, we seem to consider P to be merely ordinal, because a common substitution for P in reporting results is $-\log P$. The latter is useful for graphing purposes, as logarithmic transformations often are, and it has the further advantage that, unlike P itself, *larger* values of −logP would correspond to stronger evidence. The MLR and BF are also treated as interchangeable with their respective logarithms.[4] But logarithmic transformations of interval or ratio scaled variables create rubber scales, on which differences between measurement values are no longer meaningful.

Either we must be prepared to view ESs as (merely) ordinal, or we must accept that we are working with rubber scales. In neither case can we meaningfully compare differences, let alone ratios. But I would wager that virtually everyone – statistician, scientist and lay consumer of the scientific literature alike – interprets a change in P from, say, 0.05 to 0.04 as less exciting than a change from 0.04 to 0.001. This implies that we believe we can meaningfully compare the difference 0.05−0.04 with the difference 0.04−0.001. If we are on an ordinal or rubber scale, however, such comparisons are insupportable.

There is also widely-acknowledged incommensurability across ESs, that is, no conversion formula that would allow us to say, e.g., what value of the MLR indicates the same amount of evidence as P=0.05. Moreover, while each ES provides a rank-ordering of studies or data sets, different ESs can return different rank-orderings, raising the question of which rank-ordering is correct. Indeed, different ESs can disagree with one another regarding the simple question of which of two hypothesis is favored by a given set of data.

---

[4] Indeed the base of the logarithm is generally considered arbitrary in that it affects only the size of the degree, which is in a way true, but only as long as we are prepared to acknowledge that an ES is only ordinally scaled to begin with, in which case it was not meaningful to talk about the size of degree in the first place.

It is not difficult to see how this could be. P takes into account the probability distribution of all possible data, not just the data actually observed, assuming $H_0$. The MLR considers only the observed data, and makes an explicit comparison between two hypotheses, $H_1$ and $H_2$. The BF compares $H_1$ and $H_2$ and additionally utilizes prior probability distributions. But I am less concerned here with which of these notions of evidence, if any, is preferable, than I am with the proposition that the measure we ultimately choose should rest on a solid measurement theoretic foundation.

## 4. Absolute Scales

The designation *absolute* arises almost exclusively in connection with Kelvin's temperature scale, which we may therefore take as paradigmatic of this scale type.[5] Often the Kelvin scale is said to be absolute simply because it counts up from 0. But counting up from 0 is a feature of all ratio scales. Is there a sense in which 0°K is "absolute," while, e.g., 0 length (say, in centimeters) is not?

---

[5] In measurement theory texts the probability scale is sometimes said to be absolute in the sense that no transformations are allowable (Houle 2011), but this is a different use of the term; multiplication by a positive constant is an allowable transformation for the Kelvin scale. Also, P is on the probability scale, but insofar as it measures evidence it apparently does so merely ordinally, at best.

One way in which 0°K seems different from 0 length is that the lower bound[6] for temperature is subject to empirical determination. For instance, Amontons inferred a lower bound by extrapolating from experimental data to find the temperature in the limit as pressure went to 0, and experiments aimed at achieving ever lower temperatures are ongoing to this day. Indeed the mere existence of a lower bound on temperature was far from obvious a priori:

> "*Hot* and *cold*, like *fast* and *slow*, are mere relative terms; and, as there is no relation or proportion between motion and a state of rest, so there can be no relation between any degree of heat and absolute cold, or a total privation of heat; hence it is evident that all attempts to determine the place of *absolute cold*, on the scale of a thermometer, must be nugatory." (Rumford 1804, quoted by Chang 2004, 172)

By contrast, neither the existence of a lower bound for length nor the question of which length ought to be assigned a value of 0 requires any investigation, or even any real thought.

---

[6] In mathematical terminology, a lower bound is not a unique value, and in most occurrences in this paper a more proper nomenclature would be to refer to the greatest lower bound or infimum. But I also do not want to prejudge the issue of whether a measurement can actually =0. Therefore I will use the expression "lower bound" more colloquially, to simply refer to the bottom of the scale.

Then again, can an object have length=0?  Would it still be an object if it did? We might say that there is no such thing as a 0-length object, or perhaps there is. Nothing consequential seems to follow from a decision one way or another.[7] The situation with temperature is different. Under Kelvin's definition, 0 temperature corresponds to a fully efficient Carnot engine, and there is nothing in the theory that would prevent full efficiency from occurring; but on the other hand, the laws of thermodynamics break down at extremely cold temperatures.[8]  The most we can say is apparently Nernst's law, which tells us that for any reversible process, the T=0 isotherm cannot be intersected by any adiabat other than the S=0 isentrope (Callen 1985, 281). Never mind what that actually means. The point here is not to understand the physics, but only to note that if we want to resolve questions involving 0°K, then an understanding of physics is required.

---

[7] We might want to preclude 0 length purely on the mathematical grounds that it disrupts ratio comparisons (what is twice as much as 0?). Indeed, this seems to be part of Rumford's argument. In any case, the story turned out to be a lot more interesting for *hot* and *cold* than for *fast* and *slow* (velocity, which almost certainly *can* be 0), or length.

[8] Quantum theory precludes 0°K, but relating extremely cold quantum systems to thermodynamic theory is non-trivial in its own right.  Note also that so-called negative temperature on the Kelvin scale is a red herring: negative temperatures are actually warmer than their positive counterparts, so 0 remains the lower bound.

In short, we could say that 0°K *is* special, because it is *interesting*. The 0-point of temperature and its properties are neither obvious nor trivial to establish, but rather derive from careful study of the intended object of measurement. 0°K is *absolute* insofar as it is a lower bound established by physical laws, in a way that 0 length is not. Admittedly the distinction between 0-points for mundane ratio scales like length, and absolute minima like 0°K, is a difference of degree (no pun intended) rather than kind. Issues of measurement scale always depend on the theoretical contexts in which they arise (Houle 2011). But some theoretical contexts are more complex than others.

Thus the paradigmatic absolute scale, viz., the Kelvin scale, is simply a run-of-the-mill interval scale with lower bound of 0, with one twist: its 0-point is *absolute* in the sense of being part and parcel of the *theory of temperature* (*thermo*dynamics) in the context of which the scale is defined. It is interesting to note, however, that Kelvin's own conception of an absolute scale did not involve a 0-point at all. In fact, the first of his two temperature scales had no lower bound.  What Kelvin meant by an absolute scale was one that maintained constant meaning for the *degree* (unit) of temperature regardless of the substance being measured (Chang 2004).[9]   I return to this in the Discussion.

---

[9] Interestingly, Kelvin's second scale – the one now accepted as correct – turned out to be a linear transformation of the Celsius scale, implying that the °C "means the same" amount of temperature across the range of the scale. Therefore his first proposal, which was a non-linear transformation of Celsius, has to have been a rubber scale.

**5. Minimal Evidence**

Above I argued that P, MLS and BF are each, at best, on merely ordinal scales. Thus the existence of 0-points for these ESs is moot. Let us, however, set the issue of scale type aside for the moment and simply focus on the meaning of 0 evidence. To do so I will, like the rest of the ES-computing statistical community, play fast and loose with issues of scale, but only temporarily, in order to motivate the notion of an absolute 0-point for an ES scale.

P, by virtue of its definition as a probability, has a minimum value of 0. However, this point is interpreted as maximum, rather than minimum, evidence strength. The fact that the scale is "upside down" would not in itself be a problem for an interval scale, but we are concerned with a lower bound for a ratio scale.[10] Following standard statistical practice, we can work with $-\log P$, so that the minimum value of 0 is interpreted as the weakest possible evidence. MLR is by definition $\geq 1$ (assuming more maximization in the numerator than the denominator), that is, there is no such thing as MLR=0. As above, we can work around this problem by taking the logarithm, so that we have minimum logMLR=0, a value interpreted as the weakest possible evidence.[11]

---

[10] Some early temperature scales were oriented "upside down," with higher readings corresponding to colder temperatures (Chang 2004). On first blush, orientation seems to be merely a matter of convention. However, a ratio scale requires that the lower bound correspond to the least possible amount of the object of measurement.

The 0-points of –logP and logMLR are patently problematic. Consider a coin toss with probability $\theta$ that the coin lands heads and observed proportion $y$ of heads on $n$ tosses. For P, let $H_0$:$\theta$=½ (coin is fair); for MLR consider the hypothesis $H_1$:$\theta \neq$½ (coin is not fair) in the numerator and $H_2$:$\theta$=½ in the denominator.  If $y$=½, then –logP=0 and logMLR=0. As $n$ increases the evidence seems clearly to change (would you bet more on $\theta$=½ after $n$=2 or $n$=100?), but these ESs remain constant at 0. That is, at the 0-point, the evidence changes with $n$ but –logP=0 and logMLR=0 do not. A measure that fails to map changes in the object of measurement onto changes in scale values is inadequate in a fundamental way: it simply fails to capture the thing we set out to measure.

The 0-point of the BF is more complicated, and more interesting. The lower bound of the BF is 0, but as with P, BF=0 cannot be interpreted as a lower bound on evidence.  When BF<1, *smaller* values are said to correspond to *stronger* evidence (in favor of the denominator hypothesis $H_2$), whereas when BF>1, *larger* values are interpreted as *stronger* evidence (for $H_1$). BF=1 demarcates the boundary (or transition point, TrP) between putative evidence for one hypothesis or the other, and differences from BF=1 in either direction are taken to indicate increasing evidence. Thus BF=1 is interpreted as representing the minimal possible amount of evidence. Again, we can work with logBF so that the minimum value is 0.[12]

---

[11] Recall, however, that if P and/or MLR were on interval scales to begin with, we would now be working with rubber scales, while if –logP and logMLR were themselves on interval scales P and MLR would be on rubber scales.

LogBF is distinct from –logP and logMLR in its ability to distinguish evidence for $H_1$ from evidence for $H_2$ by virtue of the existence of a TrP, which the other ESs lack. This seems like a very nice feature of an evidence measure; even, arguably, a necessary one. However, it does complicate the issue of the 0-point. So let us first consider an artificially simple illustration.

---

[12] The domain of logBF is $(-\infty, \infty)$, rather than the positive real numbers. This in and of itself does not preclude the interpretation of logBF=0 as the minimum evidence. Houle et al. (2011) discuss the signed ratio scale, where sign is used by convention to indicate direction. They give the example of a ratio-scaled measure of left-right symmetry, where sign is used to indicate whether the symmetry is measured from left to right or from right to left; another familiar example would be physical work, which is measured on a ratio scale under the convention that positive values indicate work done by a system while negative values indicate work done to a system. If logBF constituted a ratio scale (which it does not) for evidence in favor of $H_1$, and also for evidence in favor of $H_2$, then treating logBF as a signed ratio scale would maintain the interpretation of 0 as the minimum amount of evidence, with the sign indicating "direction" (favoring $H_1$ or $H_2$). Another option is to use the absolute value of logBF, |logBF|, as I do below.

Suppose we are interested in comparing $H_1:\theta=\frac{1}{4}$ vs. $H_2:\theta=\frac{3}{4}$.[13] There are two ways we might find ourselves unable to say which hypothesis is favored: (i) we might have only irrelevant observations to go on, e.g., the number of times a die landed 3, which tells us nothing about $\theta$; or (ii) we might have the relevant observation $y=\frac{1}{2}$, which is equally compatible, and equally incompatible, with either hypothesis, because it is exactly in the middle between them. In both cases (i) and (ii), we have no basis for preferring one hypothesis over the other. But there is a distinction: we can roll the die as often as we like without affecting the evidence regarding the coin in the least; but as we continue to flip the coin another 20, 40 or 100 times, if we continue to observe $y = \frac{1}{2}$, we *are* learning something about the hypotheses, namely that the strength of the evidence against *both* of them is increasing.[14]

Here $y=\frac{1}{2}$ corresponds to the TrP of logBF: the proportion of heads such that any smaller value would tip logBF to favor of $H_1$ (logBF>0) while any larger value would tip logBF towards $H_2$ (logBF<0). The value of logBF for $y=$TrP is 0, the lower bound of the |logBF| scale, and this is true regardless of $n$ (Figure 1a). As with –logP=0 and logMLR=0,

---

[13] This gives rise to a "simple" LR (SLR), in which there are no free parameters in $H_1$ or $H_2$, and BF=SLR.

[14] Of course the inherent stochastic variability of the binomial model, and its discreteness, are incompatible with a coin landing exactly $y=\frac{1}{2}$ (or $y=$TrP, see below) on every toss. But this technicality does not undercut the gist of my argument.

interpreting |logBF|=0 as minimum evidence is therefore problematic, and in the same way: the amount of evidence we have (albeit evidence against both hypotheses) is *changing* with *n*, while |logBF| remains at its lower bound. Thus |logBF| fails to map changes in the object of measurement onto measurement changes.

**Figure 1 Transition Point (TrP) of |log BF|**

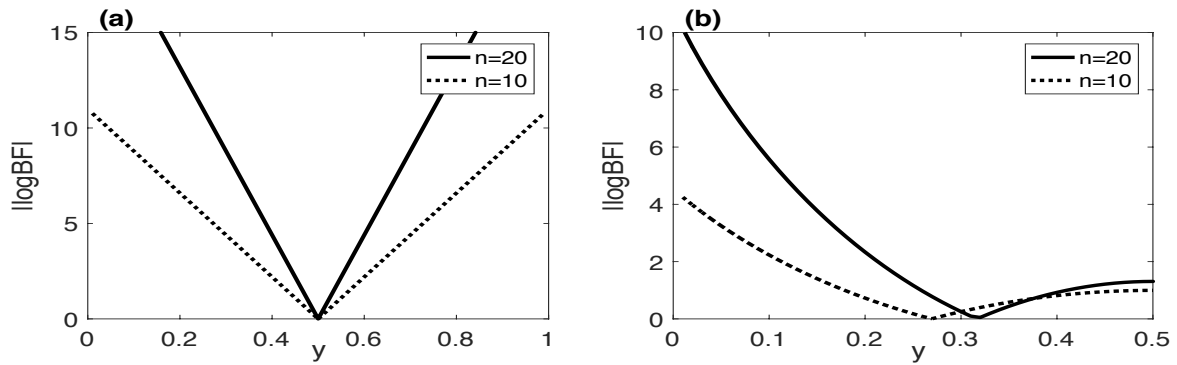

Illustrations based on the coin-tossing example from the text for (a) $H_1:\theta=\frac{1}{4}$ vs. $H_2:\theta=\frac{3}{4}$; (b) $H_1:\theta<\frac{1}{2}$ vs. $H_2:\theta=\frac{1}{2}$ (uniform prior; one-sided comparison shown for simplicity). The TrP is the point at which |logBF|=0. Values of |logBF| to the left of the TrP (logBF>0) support $H_1$, while values to the right (logBF<0) support $H_2$. In (b), the TrP moves to the right as *n* increases.

This example is special in that both hypotheses can be false (e.g., if the true value of $\theta=\frac{1}{2}$), but the reasoning generalizes. Returning now to $H_1:\theta\neq\frac{1}{2}$ vs. $H_2:\theta=\frac{1}{2}$, one or the other hypothesis must be true (with a caveat, see below). This complicates the interpretation of the TrP as representing data that are equally *incompatible* with both hypotheses, but it leaves intact the interpretation of data that are equally *compatible* with both. In addition we have a mathematical complication, because now the value of $y$ at which the TrP occurs changes with $n$ (Figure 1b). This corresponds to the familiar phenomenon of smaller effects becoming "significant" as $n$ increases.

Again it seems that the evidence increases as $n$ increases even when the data remain equally compatible with both hypotheses. But in this case, given that one or the other hypothesis must be true under the terms of the model, this would lead us to question the assumptions of the model (the aforementioned caveat). We generally assume that θ remains constant as we toss the coin (binomial model), but there is no explanation under this assumption for a coin that consistently lands in a proportion corresponding to the TrP no matter how many times we toss it, given that the TrP itself is changing. By contrast, no matter how many times we roll the die and regardless of the outcomes, we would never be led on this basis to question the binomial model for the coin. The coin toss conveys evidence, even for $y$=TrP, in a way in which rolling the die does not. Thus, as with the other ESs, there are circumstances in which the evidence is changing with $n$, but |logBF| is "stuck" at its lower bound.

## 6. Absolute Zero Evidence

If there is one thing that all statistical frameworks agree on, it is the principle that the less data we have the weaker the evidence, that is, that evidence decreases to its minimum as $n \to 0$.[15] And indeed as $n \to 0$, logBF$\to 0$ regardless of $y$. But we also have logBF($y$=TrP )=0, so another way to get to 0 is by letting $y \to$TrP.

Two descriptions of the same 0-point is not in itself a problem. For instance, there are multiple recipes for bringing a physical system to (or very close to) 0°K, e.g., by letting entropy S$\to 0$ or pressure P$\to 0$. But they all lead to the same state (viz., S≈0 and P≈0). The *meaning* of 0°K is the same regardless of how we describe the process of getting there. The situation with logBF is not like this. We can approach logBF=0 by letting $n \to 0$ or by letting $y \to$TrP, but in the latter case $n$ can be as large as we like. The two ways of getting to 0 lead to two different endpoints, or in other words, *logBF=0 is being used to mean two different things*.

What we see is that in order for the 0-point of the scale to meaningfully correspond to the limit as $n \to 0$, the ES value at the TrP *will have to change as a function of n*, so that y$\to$TrP only leads to minimal evidence when we also have n$\to 0$. The same reasoning applies, by the way, to logSLR, which also has a TrP such that logSLR($y$=TrP)=0 for all $n$. Thus it appears that logSLR=0 also does not correctly represent minimal evidence. We have known for some time that the SLR fails to provide a meaningful measure of evidence strength, even while it permits us to determine which is the better supported hypothesis (Hacking 1972). What is new here is the idea that part of the problem is something fundamentally amiss with the 0-point.

---

[15] For the binomial distribution, min($n$)=1. But the mathematical argument remains the same if we allow $n$ to continuously approach 0. A more general formulation is to say that evidence $\to 0$ as the amount of *relevant information* in the data approaches 0.

If we arrive at logBF=0 via n→0, the 0-point coincides with minimal evidence; but if we arrive there via y→TrP then, as discussed in the previous section, the 0-point does not. That |logBF|=0 does *not* always represent the weakest possible evidence is a highly counterintuitive claim, particularly as no smaller value is possible. The surprising thing is that the ES values we have always taken to represent minimal evidence cannot be pressed into service as 0-points for a proper evidence measurement scale. Insofar as there exists a lower bound on strength of evidence, and surely there is one, it begins to look like that lower bound will be *absolute*, at least in the sense of being considerably more interesting than we had anticipated.

### 7. Discussion

A great deal of science relies on an activity that looks like measurement of evidence. But useful measurement requires a cogent measurement theoretic foundation. Since we would like to be able to make meaningful evidential comparisons of order, difference and ratio, what we need is a ratio scale. I argued above that the 0-points of familiar ESs (or their variously transformed counterparts) cannot be viewed as 0-points for a proper ratio scale for evidence. I also argued that a proper ratio scale for statistical evidence measurement will need to be *absolute*, in the sense that determination of its 0-point apparently requires a better theory of evidence than what statisticians have relied on to date.

But of course a meaningful 0-point alone is not sufficient for proper measurement. An absolute scale is, at the end of the day, simply an interval scale with a lower bound of 0 that is interesting in some way, and the hallmark of an interval scale is that the unit "means the same" across the range of the scale and across contexts of application. This returns us to the concept of *absolute* in a sense closer to Kelvin's original intent. I have argued that none of the familiar ESs appear to be, or are even treated by statisticians as being, on interval scales. Of course, it is possible that one or another them happens, by sheer luck, to maintain a constant evidential meaning for the unit. But lacking a cogent theoretical foundation, this seems exceedingly unlikely and in any event impossible to substantiate.

How *does* one confirm constancy of the meaning of a unit for a theoretically constructed object of measurement? Kelvin's theory of temperature was entirely mathematical: the degree was defined in terms of ratios of heat for an ideal gas undergoing a Carnot cycle, a wholly fictional set-up that could not be implemented in the laboratory. The constancy of the meaning of the unit was embedded in the mathematics, but for that very reason, unavailable to direct empirical verification. On the other hand, by Kelvin's day there existed devices such as Amontons' air thermometer, which seemed likely, based on clever experimentation, to be measuring physical temperature on an interval scale. Thus Kelvin was able to validate his measurement scale empirically, *to some extent*, by aligning his calculations with the readings of (apparently) interval-scaled measurement devices under carefully controlled experimental conditions approximating, though never achieving, the conditions of Carnot's cycle.

An entire book could be written, however, in explication of that casual clause "to some extent" in the previous sentence (indeed, vide Chang 2004!). Constancy of the meaning of the °K as measured by actual thermometers was confirmed via a process, in Chang's phrase, of *epistemic iteration*, which to this day leaves us short of certainty, but nevertheless with a rich and productive theoretical framework. The laws of thermodynamics take on their familiar, elegant form only when expressed as a function of temperature measured on the Kelvin scale, and that is the ultimate validation of the °K. But it remains an unassailable fact that there is no such thing as direct verification that any given measurement device is consistently measuring on the Kelvin scale, let alone doing so under all conditions of application.

Kelvin had something else working in his favor, apart from reasonably good thermometers: he was among a community of scientists with a shared desire for a better understanding of temperature. By contrast, it seems impossible to interest statisticians in the need for a better understanding of evidence. Perhaps this is because they view the very idea of measurement of evidence on a proper scale to be, in a word, nugatory. After all, how would we verify that one degree of evidence on any given measurement scale always "means the same" with respect to the evidence, without some independent way of knowing what the evidence is?

The point is well taken, but moot. Vindication of a theoretical measurement construct is not a matter of axiomatics. It happens by epistemic iteration, not in one fell swoop and never to the point of mathematical certainty. Perhaps the first step to solving the evidence measurement problem is understanding the limits on what demonstration of a solution would look like.

**Acknowledgments** This work was supported by a grant from the W.M. Keck Foundation. I thank Susan E. Hodge for helpful comments on an earlier draft, and the faculty of the Battelle Center for Mathematical Medicine for allowing me to use them as a "focus group" on the meaning of absolute 0 temperature. Many of the ideas in the paper emerged from ongoing discussions with my collaborator, Sang-Cheol Seok, without whom our evidence measurement project would have long since come to a standstill.